\documentclass[seceq,epsf,preprint]{ptptex}
\usepackage[xdvi,dvips]{graphicx}

\newcommand{\la}{\langle}
\newcommand{\ra}{\rangle}

\newcommand{\beq}{\begin{eqnarray}}
\newcommand{\eeq}{\end{eqnarray}}
\newcommand{\sbeq}{\begin{subeqnarray}}
\newcommand{\seeq}{\end{subeqnarray}}
\renewcommand{\theequation}{\thesection.\arabic{equation}}

\newcommand{\btem}{\bibitem}

\newcommand{\bfm}{\mbox{{\boldmath $m$}}}

\preprintnumber[5cm]{
KUNS-1809, YITP-02-65}

\title{%
Chiral and Color-superconducting 
Phase Transitions with Vector Interaction in a Simple Model
}

\author{
Masakiyo
\textsc{Kitazawa}$^a$\thanks{masky@ruby.scphys.kyoto-u.ac.jp}, 
Tomoi \textsc{Koide}$^b$\thanks{tkoide@yukawa.kyoto-u.ac.jp}, 
Teiji \textsc{Kunihiro}$^b$\thanks{kunihiro@yukawa.kyoto-u.ac.jp} 
and Yukio \textsc{Nemoto}$^c$\thanks{nemoto@bnl.gov}
}

\inst{%
$^a$ Department of Physics, Kyoto university, Kyoto 606-8502, Japan \\
$^b$ Yukawa Institute for Theoretical Physics, Kyoto University, Kyoto 606-8502, Japan \\
$^c$ RIKEN BNL Research Center, BNL, Upton, NY 11973
}

\abst{
We investigate effects of the 
vector interaction on chiral and color superconducting (CSC) 
phase transitions at finite density and temperature in a simple 
Nambu-Jona-Lasinio model.
It is  shown that the repulsive density-density interaction 
coming from the vector term, which is present in the effective
chiral models but has been omitted, enhances 
the competition between the chiral symmerty breaking ($\chi$SB)
and CSC phase transition, and thereby makes 
the thermodynamic potential have a shallow minimum
over a wide range of values of the correlated 
chiral and CSC order parameters.
We find that when the vector coupling is increased,
the first order transition
between the $\chi$SB and CSC phases becomes weaker, 
and the coexisting phase
in which both the chiral and
color-gauge symmetry are dynamically broken comes 
to exisit over a wider range of the density and temperature.
We also show that there can exist two endpoints, which
are tricritical points in the chiral limit, 
along the critical line of the 
first order transition in some range of values of the vector coupling.
Although our analysis is based on a simple model,
the nontrivial interplay between the $\chi$SB and CSC phases
induced by the vector interaction 
is expected to be a universal phenomenon 
and might give a clue to understanding results obtained with 
two-color QCD on the lattice.
}
\begin{document}

\maketitle

\section{Introduction}

It is one of the central issues in hadron physics
to determine the  phase diagram of 
strongly interacting matter  in the temperature ($T$)-chemical 
potential ($\mu$) 
or $T$-$\rho_B$ plane, with $\rho_B$ being the baryonic density.
In extremely hot and dense matter,
the non-Abelian nature of QCD  ensures 
that the colored quarks and gluons are not confined, and 
chiral symmetry is restored.
Lattice simulations of QCD \cite{ref:Lattice,ref:Lattice2} show that 
the QCD vacuum undergoes a chiral and deconfinement transition 
at a temperature $T_c$ around 150$-$175 MeV at vanishing
chemical potential,
with the order and critical temperature being 
dependent on the number of active flavors.
Although there have been several promising attempts
\cite{ref:FK,ref:Naka,ref:MNN,ref:KSHM,ref:KTS,ref:KTS2,ref:AACGG} 
to make simulations of lattice QCD with finite $\mu$ possible,  
they have still not progressed enough to predict anything definite 
about the phase transition at finite $\rho_B$ or $\mu$.
It is widely believed on the basis of 
effective theories\cite{ref:Kle,ref:HK,ref:SS} and 
chiral random matrix theory\cite{ref:VW} that 
the chiral phase transition 
from the chiral-symmetry broken to the restored phase 
is first order at vanishing temperature.
Furthermore, people believes now that 
the critical line of the first order chiral transition 
continues for smaller chemical potentials in the $T$-$\mu$ plane
and ends at some point with $T=T_e$ and $\mu =\mu _e$, which is
called the endpoint.  
We notice that the first-order chiral transition is
accompanied by a jump in the baryon density.

Recent renewed interest in color superconductivity (CS)
\cite{ref:Barr,ref:BL,ref:Iwa,ref:ARW,ref:RSSV,ref:PR,ref:Son,ref:SW,ref:PR2} 
has stimulated intensive studies of the 
QCD phase structure at finite density in the low temperature
region, which in turn are revealing a rich phase structure of 
high density hadron/quark matter with CS
\cite{ref:BR,ref:RSSV-coex,ref:CFL,ref:ABR,ref:BO,ref:GNA,ref:HT}.
Possible relevance of CS to characteristic phenomena observed
for neutron stars are being actively discussed
\cite{ref:RW-review,ref:A}.
Some recent studies have also suggested that experiments on the Earth 
using heavy-ion collisions with large baryon stopping 
can elucidate something about CS in dense matter\cite{ref:KKKN,ref:PR3}.

The purpose of the present paper is to reveal new characteristics 
of the chiral to color superconducting (CSC) transition based on 
a simple effective model incorporating the vector interaction 
by focusing on the implication of the density jump accompanied 
by the chiral transition.

Low-energy effective models\cite{ref:Kle,ref:HK,ref:SS,ref:BR,ref:RSSV-coex}
are useful 
to study not only the chiral transition but also CS 
in  dense hadronic matter.
For example, chiral models of the Nambu-Jona-Lasinio type, 
\cite{ref:NJL} which can be considered simplified versions 
of those with an instanton-induced interaction, accurately describe 
the gross features of the $T$ dependence of the chiral 
quark condensates of the lightest three quarks as given 
by lattice QCD, and predict that the chiral transition for $\mu\not=0$
is rather strongly first order at low 
temperatures when the vector interaction is absent or small 
\cite{ref:AY,ref:Kuni,ref:SS}.
Chiral effective theories 
show that the gap $\Delta$ of CS may become 
as large as 100 MeV in relatively low densities, where 
a phase change from the chiral symmetry breaking ($\chi$SB) 
phase to the CSC phase may also occur\cite{ref:BR,ref:RSSV-coex}.

However, although  many works 
on CS have been carried out with the use of effective models, 
the vector interaction
\cite{ref:ES,ref:Klei,ref:ER,ref:Volk,
ref:BMZ,ref:AY,ref:KLW,ref:LKW,ref:Kuni2,ref:EKMV,ref:Bub,
ref:LR,ref:BHO},
\beq
{\cal L}_V=-G_V (\bar{\psi} \gamma ^{\mu}\psi)^2,
\label{lv}
\eeq
has been scarcely taken into account, 
with the exception of very recent works
\cite{ref:LR,ref:BHO}.
Our point is that such a vector interaction is 
chiral invariant and naturally 
appears in the effective models derived from microscopic
theories and, as we shall show, indeed 
can have strong effects on the chiral-to-CSC transition 
and the properties of the CSC phase.

Although it may not be a common knowledge 
in the physics community, 
the importance of the vector coupling for the chiral transition 
is known; i.e., 
the vector coupling weakens the phase 
transition and moves the chiral restoration to a larger value of $\mu$
\cite{ref:AY,ref:KLW,ref:Bub}.
This can be intuitively understood as follows\cite{ref:kuni_conf}.
According to thermodynamics, when  two phases I and II 
are in an equilibrium state, 
their temperatures $T_{\rm I, II}$, pressures $P_{\rm I, II}$ and 
the chemical potentials $\mu_{\rm I, II}$ are the same:
\beq
T_{\rm I}=T_{\rm II}, \quad \  
P_{\rm I}=P_{\rm II}, \quad \  
\mu_{\rm I}=\mu_{\rm II}.
\eeq
If I and II are the chirally broken and  restored phase 
with quark masses satisfying $M_I>M_{II}$, 
the last equality further tells us that
the chirally restored phase has a higher density than the
broken phase, because $\mu_{\rm I, II}$ 
at vanishing temperature are given 
by $\mu_{\rm i}=\sqrt{M_i^2+p_{F_i}^2}$, ($i=$ I, II),
and hence $p_{F_I}<p_{F_{II}}$, 
where  $p_{F_i}$ is the Fermi momentum of the $i$-th phase.
Thus it is seen that chiral restoration at finite density is necessarily
accompanied by a density jump to a higher density state with a
large Fermi surface, which in turn favors the formation of Cooper
instability leading to CS.

However, since the vector coupling includes the term 
$(\bar{\psi}\gamma^0 \psi)^2$, 
it gives rise to a repulsive energy proportional to the density squared, i.e.
$G_{V}\rho_B^2/2$, which is larger in the restored phase than in the 
broken phase; the vector coupling weakens and delays the phase 
transition of the chiral restoration at low temperatures.
Thus one expects naturally that ${\cal L}_V$
causes the chiral restoration and the formation of CS 
to shift to higher chemical potentials,
and may alter the nature of the transition 
from the $\chi$SB phase to the CSC phase drastically.

Is it legitimate to include a vector term like 
(\ref{lv}) in an effective Lagrangian?
First of all, one should notice that 
the instanton-anti-instanton molecule model,\cite{ref:SS,ref:RSSV-coex} as well 
as the renormalization-group equation,\cite{ref:EHS,ref:SW-reno} shows that 
${\cal L}_V$ appears as a part of the effective 
interactions together with those in the scalar channels, 
which are responsible for the chiral symmetry breaking ($\chi$SB):
The instanton-anti-instanton molecule model 
gives for the effective interaction between quarks 
\beq
{\cal L}_{mol sym}&=&G_{mol}\bigl\{
\frac{2}{N_c^2}\left[(\bar{\psi}\tau^a\psi)^2+
(\bar{\psi}\tau^ai\gamma _{5}\psi)^2\right]\nonumber \\
 &  &-\frac{1}{2N_c^2}\left[(\bar{\psi}\tau^a\gamma _{\mu}\psi)^2-
(\bar{\psi}\tau^a\gamma _{\mu}\gamma_5\psi)^2\right]
+\frac{2}{N_c^2}(\bar{\psi}\gamma _{\mu}\gamma_5\psi)^2\bigl\}+
{\cal L}_8,
\eeq
where $\tau^a=(\vec{\tau}, 1)$ and ${\cal L}_8$ denotes 
the color octet part of the interaction, 
which we shall not write down.
Near the phase transition,
the instanton molecules are polarized 
in the temporal direction, Lorenz invariance is broken, and 
thus the vector interactions are modified as 
$(\bar{\psi}\gamma_{\mu}\Gamma\psi)^2 \rightarrow
 (\bar{\psi}\gamma_{0}\Gamma\psi)^2$.
Notice that 
the instanton-induced interaction breaks the U$_A(1)$ symmetry.
In reality, however, there should also exist 
U$_A(1)$-symmetric interactions such as the one-gluon 
exchange interaction or its low-energy remnant as 
\beq
{\cal L}^0_{LL}
=G^0_{ll}\left\{(\bar{\psi}_L\gamma_0\psi_L)^2
-(\bar{\psi}_L\gamma_i\psi_L)^2\right\},
\eeq
where $\psi_L$ denotes the left-handed quark field.
It is shown using the renormalization group equation 
that the strengths of the U$_A(1)$-symmetric 
and violating effective interactions are of the same order near 
the Fermi surface.
Thus one sees that the vector interaction exists 
together with other chiral invariant terms which are usually used.
Therefore, 
one may say that the previous works dealing with the
$\chi$SB-to-CSC phase transition  without incorporating 
the vector interaction ${\cal L}_V$ are all incomplete,
because this interaction may alter the nature of the phase transition 
significantly.

We shall show in this paper that
the inclusion of the vector coupling 
induces a novel interplay between the $\chi$SB and CS through 
the difference of the respective favoring baryon densities and changes 
both the nature of the phase transition and the phase structure
in the low temperature region drastically
\footnote{Preliminary results have been reported in Ref. \citen{ref:JPS}}.
The resultant phase diagram 
and the behavior of the chiral and diquark condensates as functions 
of $(T, \mu)$ will be found to have a good correspondence with 
those given in two-color QCD on the lattice.\cite{ref:KTS2}
It is thus found that our simple model
gives a possible mechanism underlying the lattice results.

This paper is organized as follows.
In the next section, the Lagrangian to be used is introduced.
In \S 3, we shall give the thermodynamic
potential and the self-consistency condition for the 
quark condensate and the pairing field.
Numerical results are presented in \S4.
The final section is devoted to a summary and concluding remarks.
The appendix summarizes the effects of the vector interaction 
on the chiral transition when the CS is not incorporated.

\section{Model}

As a chiral effective model which embodies the vector interaction as well
as the usual scalar terms driving $\chi$SB,
we use a simple
Nambu-Jona-Lasinio (NJL) model 
with two flavors ($N_f=2$) and three colors ($N_c=3$),
following Ref. \citen{ref:BR}.
The NJL model may be regarded as a simplified version of that with 
instanton-induced interactions and 
can also be derived using a Fierz transformation 
of the one-gluon exchange interaction
with heavy-gluon approximation (see
\citen{ref:Kle,ref:HK,ref:Vogl,ref:ERV,ref:Alk}).
This effective model has the merit that it can be used to investigate 
the chiral transition and CS simultaneously, and hence describes 
their interplay.
It was shown \cite{ref:SKP} 
that the physical content given with the instanton model\cite{ref:BR} 
can be nicely reproduced by the simple NJL model
with a simple three-momentum cutoff.
This means that although there are several choices for the high momentum 
cutoff which mimics the  asymptotic freedom, 
the magnitude of the gap is largely determined by the strength 
of the interaction and is insensitive to the form of the momentum cutoff 
\cite{ref:RW-review}.
The Lagrangian density thus reads
\beq
{\cal L}= {\cal L}_0+{\cal L}_I, 
\eeq
where 
\beq
{\cal L}_0=\bar{\psi}(i\gamma\cdot \partial -\bfm)\psi,
\eeq
with $\bfm$ being the current quark mass matrix
$\bfm ={\rm diag}(m_u, m_d)$,
and 
\beq
{\cal L}_I = {\cal L}_S+{\cal L}_V+{\cal L}_C,
\label{eqn:NJL}
\eeq
with
\beq
{\cal L}_S=G_S \big\{ (\bar{\psi}\psi)^2
+(\bar{\psi} i \gamma_5 {\bf \tau } \psi)^2 \big\},
\eeq
and
\beq
{\cal L}_C=
 G_{C} \big\{ (\bar{\psi} i \gamma_5 \tau_2 \lambda_A \psi^C )
(\bar{\psi}^C i \gamma_5 \tau_2 \lambda_A \psi)
+ (\bar{\psi} \tau_2 \lambda_A \psi^C )
(\bar{\psi}^C \tau_2 \lambda_A \psi) \big\}.
\eeq
${\cal L}_V$ is given in (\ref{lv}).
Here, $\psi^{C}\equiv C\bar{\psi}^{T}$, with 
$C=i\gamma^{2}\gamma^{0}$ 
being the charge conjugation operator, and 
$\tau_{2}$ and $\lambda_{A}$'s are the second component 
of the Pauli matrix representing the flavor SU$(2)_{f}$, 
and the antisymmetric Gell-Mann matrices 
representing the color SU$(3)_c$, respectively.
The scalar coupling constant $G_S=5.5~{\rm GeV}^{-2}$ 
and the three momentum cutoff $\Lambda = 631$ MeV are chosen 
so as to reproduce the pion mass $m_{\pi}=139$ MeV and the pion 
decay constant $f_{\pi}=93$ MeV with 
the current quark mass $m_u=m_d=5.5$ MeV\cite{ref:HK};
we have assumed isospin symmetry. 
It should be noted that the existence of the diquark coupling 
$G_{C}$ and the vector coupling $G_{V}$ 
do not affect the determination of the pion decay constant 
and the chiral condensate.
Although, there are several sources to determine 
the diquark coupling such as
the diquark-quark picture of baryons\cite{ref:Vogl,ref:ERV,ref:Alk,ref:MBIY}, 
the instanton-induced interaction\cite{ref:RSSV-coex}, 
renormalization group analysis,\cite{ref:EHS,ref:SW-reno} and so on,
we shall take $G_C/G_S=0.6$, which 
accurately reproduces the phase diagram obtained with 
the instanton-induced interaction\cite{ref:BR}.
As for the vector coupling,
we vary it as a free 
parameter in the range of $G_{V}/G_{S} = 0 - 0.5$
to see the effect of the vector coupling on the phase diagram.
We remark that the vector coupling $G_{V}$ is given by $0.25G_{S}$ 
in the instanton-anti-instanton molecule model\cite{ref:RSSV} 
and $0.5G_{C}$ in the renormalization-group
analysis\cite{ref:EHS,ref:SW-reno};
the range we employ for $G_V$ thus encompasses these {\em physical}
values.

\section{Thermodynamic Potential and Gap Equations}

In this section, we calculate the thermodynamic potential in the 
mean-field approximation 
and derive 
the coupled gap equations for the chiral and diquark condensates.

The thermodynamic potential $\Omega$ 
is defined by 
\begin{eqnarray}
\Omega = -T\ln~{\rm Tr}~{\rm e}^{-\beta \hat{K}}, \label{ref:TP}
\end{eqnarray}
where $\beta=1/T$ is the inverse temperature and 
$\hat{K}=\hat{H}-\mu \hat{N}$, with $\hat{H}$ and 
$\hat{N}$ being the Hamiltonian and the
quark number operator, respectively. 
The expectation value of the quark number is given by
\beq
N_q=\la\hat{N}\ra,
\eeq
where 
\beq
\la \hat{\cal O}\ra=
{\rm Tr}~{\rm e}^{-\beta (\hat{K} -\Omega)}\hat{\cal O}
\eeq
denotes the statistical average of $\hat{\cal O}$.
The quark number density is given by
\beq
\label{def-rho}
\rho_q=N_q/V=
\langle \bar{\psi}\gamma^{0}\psi \rangle,
\eeq
where $V$ denotes the volume of the system, 
and it is assumed that 
the vacuum contribution to the quark number is subtracted.\cite{ref:BD}
\footnote{
The rotational invariance of the system, which we assume, 
implies that the spatial component of the expectation value 
$\langle \bar{\psi}\gamma^{i}\psi \rangle$ vanishes.}

The quark number $N_q$ can be calculated 
by means of a thermodynamic relation from $\Omega$ as
$N_q =-\partial \Omega/\partial \mu$, and accordingly,
$\rho_q$ is obtained from the thermodynamic
potential density as\footnote{This is a familiar procedure 
in the $\sigma$-$\omega$ model\cite{ref:SigOmeg}.}
\begin{eqnarray}
\rho_q=-\frac{\partial (\Omega/V)}{\partial \mu}. \label{eqn:TR}
\end{eqnarray}

Since quarks have baryon number $1/3$, 
the baryon number density and chemical potential are given by 
$\rho_B=1/3\cdot \rho_q$ and 
$\mu_B=3\mu$, respectively, where iso-spin symmetry is assumed.
We shall use the quark number
density $\rho_q$ and chemical potential $\mu$ for the formulation,
but $\rho_B$ and $\mu_B$ will be used in the presentation
of the numerical results in \S 4.

To apply the mean-field approximation (MFA),
we first assume that 
the system has a quark-antiquark condensate
$\langle \bar{\psi} \psi \rangle$
and a diquark condensate 
$\langle\bar{\psi}^C i\gamma_5 \tau_2 \lambda_2 \psi \rangle$,
where $\lambda_{A}$ is restricted to $\lambda_2$ 
owing to the color SU$(3)_{c}$ symmetry. 
In the MFA, $\hat{K}$ is replaced by  
\begin{eqnarray}
\hat{K}_{\rm MFA}
&=& \int d^3 {\bf x}\left[ \bar{\psi}[-i\vec{\gamma}\cdot \vec{\nabla }
+(m+M_{D})-( \mu - 2G_{V}\rho_q ) \gamma_{0}]\psi
+\frac{1}{2}
(\Delta^{*}\bar{\psi}^{C}i\gamma_{5}\tau_{2}\lambda_{2}\psi 
+ {\rm h.~c.})
\right. \nonumber \\
&&\left. +\frac{M_{D}^2}{4G_{S}}+\frac{|\Delta|^2}{4G_{C}}- G_{V}
\rho_q^2 \right].\label{eqn:FE}
\end{eqnarray}
Here, $M_D$ and $\Delta$ give the dynamically generated quark mass 
and the gap due to the CS, respectively:
\begin{eqnarray}
M_{D} =  - 2 G_S \langle \bar{\psi} \psi \rangle, \qquad
\Delta = -2 G_{C} \langle
\bar{\psi}^C i\gamma_5 \tau_2 \lambda_2 \psi \rangle.
\end{eqnarray}
We notice here that  
$\mu$ in $\hat{K}_{\rm MFA}$ appears in the combination 
\begin{eqnarray}
 \mu-2G_{V}\rho_q\equiv \tilde{\mu}. 
\label{eqn:tilmu}
\end{eqnarray}
Thus, the thermodynamic potential 
$\Omega _{\rm MFA}$ in MFA per unit volume is calculated 
to be
\begin{eqnarray}
\omega(M_{D},\Delta;T,\mu) 
&\equiv& {\Omega _{\rm MFA}}/{V} \nonumber \\
&=& \frac{ M_{D}^2 }{ 4G_S } 
+ \frac{ |\Delta|^2 }{ 4G_C } - G_V \rho_q^2
\nonumber \\
&& -4 \int \frac{ d^3 p }{ (2\pi)^3 }
\left\{ E_{\bf p} + T \log \left( 1 + e^{ -\beta \xi_- } \right)
\left( 1 + e^{ -\beta \xi_+ } \right) \right.\nonumber \\
&&\left. + sgn(\xi_{-})~\epsilon_- + \epsilon_+ + 2T \log
\left( 1 + e^{ -sgn(\xi_{-})\beta \epsilon_- } \right) \left( 1 + e^{ -\beta \epsilon_+ }
\right) \right\},\nonumber \\
\label{eqn:2omega}
\end{eqnarray}
where 
$E_{p} = \sqrt{p^2 + M^2}$, 
$\xi_{\pm} = E_{p} \pm \tilde{\mu}$ and 
$\epsilon_{\pm} = \sqrt{\xi_{\pm}^2 + |\Delta|^2}$,
with 
\beq
M=m+M_{D} \label{conmass}
\eeq
being the total (constituent) quark mass and $sgn(\xi_{-})$ 
the sign function.
Our thermodynamic potential reduces to those given in 
Refs.~\citen{ref:BR, ref:SKP} when $G_V=0$. 
The quark density $\rho_q$ appearing in (\ref{eqn:2omega})
is expressed as a function of
the condensates $(M_D, \Delta)$
through  the thermodynamical relation (\ref{eqn:TR})
as 
\begin{eqnarray}
\rho _q
= 4 \int \frac{ d^3 p }{ (2\pi)^3 }
\left\{ n ( \xi_- ) - n ( \xi_+ ) - \frac{ \xi_- }{ \epsilon_- }
\tanh \frac{ \beta \epsilon_- }2
+ \frac{ \xi_+ }{ \epsilon_+ } \tanh \frac{ \beta \epsilon_- }2
\right\},
\label{eqn:2GER}
\end{eqnarray}
where $n(\xi_{\pm})$ is the Fermi distribution function: 
$n(\xi_{\pm})=1/(\exp\{\beta \xi_{\pm} +1\})$.
Equation (\ref{eqn:2omega}) together with Eq.(\ref{eqn:2GER})
gives the thermodynamic potential $\omega$ with  the
condensates $(M_D, \Delta)$ being  the variational parameters at 
given $(T, \mu)$; 
their optimal values give the absolute
minimum of $\omega$.

The chiral and diquark condensates in the equilibrium state
at given $(T, \mu)$
should satisfy the stationary conditions
 for the thermodynamic potential,
\begin{eqnarray}
\frac{\partial \omega}{\partial M_{D}}\Bigl\vert_{\Delta} =0,~~~
\frac{\partial \omega}{\partial \Delta}\Bigl\vert_{M_D} = 0, 
\end{eqnarray}
which are reduced to the self-consistency conditions for 
the two condensates,
\begin{eqnarray}
M_D &=& 8 G_S M \int \frac{ d^3 p }{ (2\pi)^3 }
\frac1{ E_p } \left\{ 1 - n ( \xi_- ) - n ( \xi_+ )
+ \frac{ \xi_- }{ \epsilon_- } \tanh \frac{ \beta\epsilon_- }2
+ \frac{ \xi_+ }{ \epsilon_+ } \tanh \frac{ \beta\epsilon_+ }2 \right\},
\label{eqn:2GEM}\nonumber  \\
\\
\Delta &=& 8 G_{C} \Delta \int \frac{ d^3 p }{ (2\pi)^3 }
\left\{ \frac1{ \epsilon_- } \tanh \frac{ \beta\epsilon_- }2
+ \frac1{ \epsilon_+ } \tanh \frac{ \beta\epsilon_+ }2 \right\}.
\label{eqn:2GED} 
\end{eqnarray}
Here we have utilized the chain rule 
\beq
\frac{\partial \omega}{\partial M_{D}}\Bigl\vert_{\Delta} =
\frac{\partial \omega}{\partial M_{D}}\Bigl\vert_{\Delta, \rho_q}+
\frac{\partial \rho_q}{\partial M_D}\Bigl\vert_{\Delta}\cdot 
\frac{\partial \omega}{\partial \rho_q}\Bigl\vert_{M_D, \Delta,},
\eeq
and that for the $\Delta$-derivative, 
together with the fact that Eq. (\ref{eqn:2GER}) ensures the relation
\begin{eqnarray}
\frac{\partial \omega}{\partial \rho_q}\Bigl\vert_{M_D, \Delta} =0.
\label{eqn:dodr}
\end{eqnarray}
In analogy to the BCS theory of the superconductivity, 
we call Eqs. (\ref{eqn:2GEM}) and (\ref{eqn:2GED}) the gap 
equations. Notice, however, that 
a solution of the gap equations may only give
a {\em local} minimum, or even {\em maximum} of the 
thermodynamic potential, and
it is  only a candidate of the optimal value of the condensates;
one must check whether it gives the absolute minimum of 
the thermodynamic potential.

From the structure of the coupled gap equations and the
thermodynamic relation (\ref{eqn:2GER}) for $\rho_q$,
one can extract some interesting properties of the
condensates $(M_D, \Delta)$ as functions of $(T, \mu)$
and also of $(T, \rho _q)$.

(1)~
Once the absolute minimum of the thermodynamic potential 
and, accordingly, the optimal value
of $(M_D, \Delta)$ are found at given $(T, \mu)$, 
the quark density $\rho_q$ is given by  Eq.~(\ref{eqn:2GER}).
The coupled gap equations (\ref{eqn:2GEM}) and (\ref{eqn:2GED})
show us  that the optimal value of $(M_{D}, \Delta )$ is a 
function of  $T$ and $\tilde{\mu}$, and in this way the 
possible $G_{V}$ dependence is absorbed into $\tilde{\mu}$.
Furthermore, if $(M_D(T, \mu), \Delta (T, \mu))$
is a solution of the coupled gap equations with $G_V=0$,
then 
\beq
(M_D(T, \tilde{\mu}), \Delta (T, \tilde{\mu}))\equiv
(M_D(T, \mu -2G_V\rho_q), \Delta (T, \mu-2G_V\rho_q))
\label{shift}
\eeq
is a solution with $G_V\not=0$. Thus,
the whole solution as a function 
of $\mu$ is shifted toward larger $\mu$ by an amount $2G_V\rho_q$.

(2)~ Next, we shall examine how the solutions of the 
coupled gap equations behave as functions of $(T, \rho_q)$ 
instead of $(T, \mu)$.
Let the $0$-th order approximation of the 
condensates be given.
Then Eq. (\ref{eqn:2GER}) gives $\tilde{\mu}$  as a function
of $(T, \rho_q)$, i.e., $\tilde{\mu}=\tilde{\mu}(T, \rho_q)$.
Thus, the first-order approximation of the condensates 
$(M_{D}, \Delta )$ is given as the solution to
the coupled gap equations (\ref{eqn:2GEM}) and (\ref{eqn:2GED}), 
which are only dependent on $T$ and $\tilde{\mu}$, 
not on $\mu$ and $\rho_q$, separately;
possible $G_{V}$ dependence is absorbed into $\tilde{\mu}(T, \rho_q)$.
Thus, repeating this procedure,
one sees that $(M_{D}, \Delta)$ becomes only a 
function of $T$ and $\rho_q$ and is independent of 
$G_V$, because $\tilde{\mu}$, through which 
$G_V$ can affect the formulas, actually only plays 
the role of a dummy variable.
Thus we have proved that there is
no effect of the vector interaction on 
the behavior of the solution to the coupled gap equation as 
functions of $(T, \rho_q)$.
\begin{figure}
\begin{center}
{\includegraphics[scale=1]{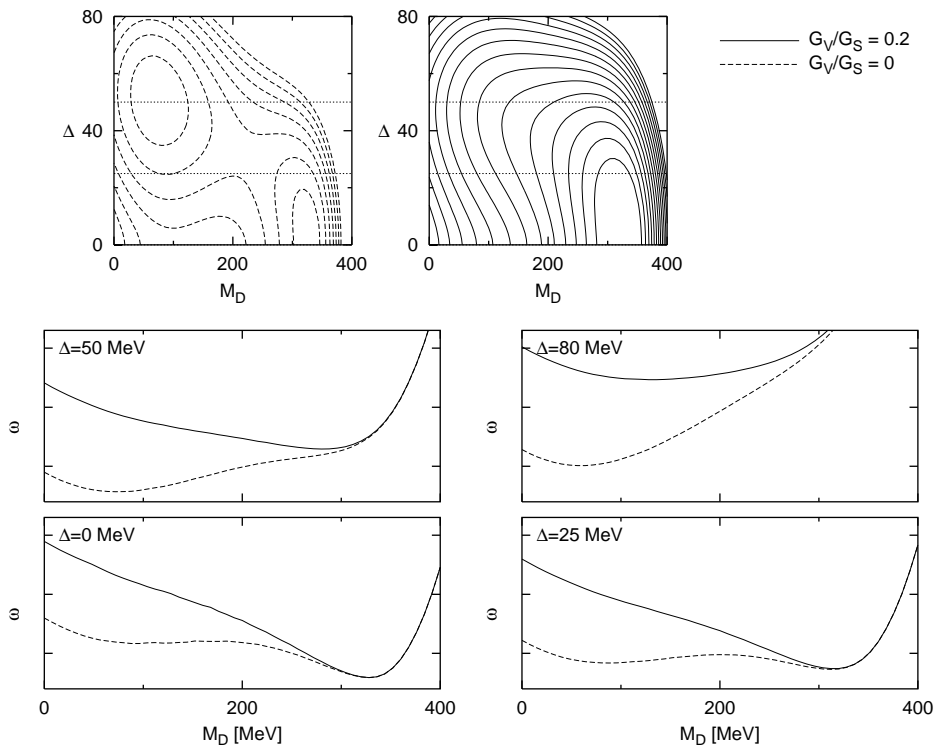}}
\caption{The upper panels show contour maps
 of the thermodynamic potential $\omega$ 
at $(T, \mu_B)=(0, 1035)$ MeV with $G_V/G_S=0$
(left panel) and $G_V/G_S=0.2$ (right panel) 
in the $M_D$-$\Delta$ plane.
The difference between the values of $\omega$ for the adjacent 
contour lines is $7.5\times 10^6 \mbox~{MeV}^4$.
The lower four panels are
cross sections of $\omega$ cut in a plane with $\Delta$ fixed
 at $0, 25, 50$ and $80$ MeV. The solid (dashed) lines 
represent the $G_V/G_S=0.2$\, ($G_V/G_S=0$) case.
}
\label{fig:ep3}
\end{center}
\end{figure}

(3)~ Does it mean that there is no trace of the presence of the
vector interaction in the phase diagram in the 
$T$-$\rho$ plane?
The answer is no.
The effect of the vector interaction
manifests itself in the critical point or line when
 the transition is first order.
In this case, there are several solutions to the coupled gap equations 
(\ref{eqn:2GEM}) and (\ref{eqn:2GED}),
corresponding to the local minima, maxima and even saddle points 
of $\omega$;
notice that these solutions correspond to different baryon densities.
Since  the thermodynamic potential (\ref{eqn:2omega}) is explicitly
dependent on $G_{V}$ in a combination with the quark density, 
the relative magnitudes of the local minima
change and can be altered with the vector interaction:
In Fig. \ref{fig:ep3}, the right (left) figure in the upper panel 
shows  the contour map of the 
thermodynamic potential $\omega(M_D, \Delta)$ 
with $G_V/G_S=0.2$ ($G_V/G_S=0$) at $T=0$ and $\mu_B=\mu_{B0}=1035$ MeV,
which is actually found to be the critical point.
The thermodynamic potential
$\omega(M_D, \Delta)$ as a function of
$M_D$ at given $\Delta=0, 25, 50$ and $80$ MeV; i.e.,
the cross sections along the  lines shown in the upper panels 
are given in the lower panels, where 
the solid (dashed) lines denote $\omega$ 
with $G_V/G_S=0.2$ ($G_V/G_S=0.$).  
One clearly sees that the vector interaction increases the
thermodynamic potential in the small $M_D$ region for every $\Delta$;
notice that the system with smaller $M_D$ is at higher density, 
as discussed in \S1. 
Thus the absolute minimum given with $\Delta\sim 50$ MeV at small 
$M_D$ when $G_V=0$ ceases to be even a local minimum with 
finite $G_V/G_S$, and the local minimum at $M_D\sim 300$ MeV with 
$\Delta\sim 0$ in turn becomes the unique local, and hence, 
the absolute minimum.
Thereby the double-minimum structure disappears, and 
the first order transition is altered to a crossover.
In short, the critical temperatures and densities at which 
the transition from one local minimum to the other occurs 
are strongly affected by the vector interaction, and 
the critical line of the first-order transition in 
the $T$-$\rho_q$ plane is  changed with the vector interaction.

In passing, we remark 
that $\rho_q$ cannot be interpreted as a variational parameter 
with which the thermodynamic potential is minimized:
Since Eq.~(\ref{eqn:2GER}) is obtained by the stationary 
condition Eq. (\ref{eqn:dodr}), 
one might have imagined that the equilibrium state could be determined by 
searching for the minimum point of the thermodynamic
potential with $\rho_q$ being a variational parameter
together with $M_{D}$ and 
$\Delta$\cite{ref:LR}.
However, 
Eq.~(\ref{eqn:dodr}) is found to give a local {\em maximum} 
of the thermodynamic potential.
That is, the absolute minimum of the thermodynamic
potential in the $M$-$\Delta$-$\rho_q$ space, if it exists, 
does not give the thermodynamical equilibrium state.

\section{Numerical results and discussion}

In this section, we show the numerical results and
discuss the effects of the vector coupling on the 
phase diagram, the $T$-$\mu_B$  and
$T$-$\rho_B$ dependence of the order parameters.

\subsection{Phase diagram with no vector interaction}

\begin{figure}[t]
\begin{center}
\begin{tabular}{cc}
{\includegraphics[scale=1]{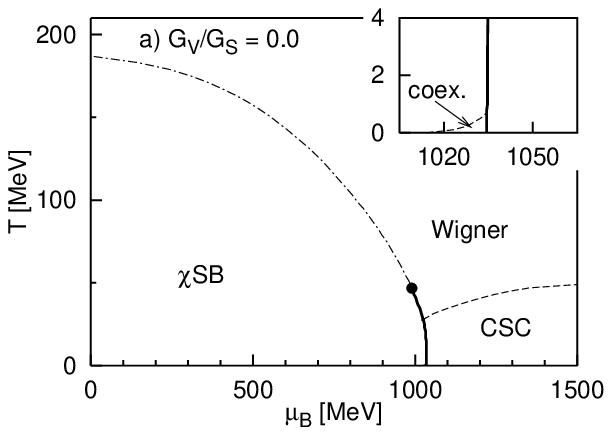}} & \hspace{-5mm}
{\includegraphics[scale=1]{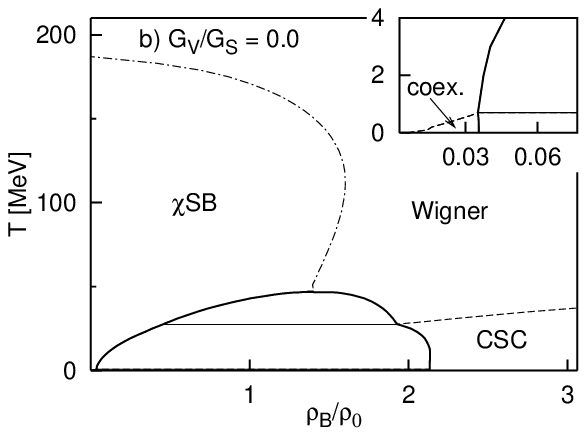}} \\
\end{tabular}
\caption{
(a)~The phase diagram in the $T$-$\mu_B$ plane with $ G_V=0 $.
There are four phases; the $\chi$SB, CSC, Wigner, and coexisting phases.
The small panel is an enlargement
around the border of the $\chi$SB and CSC phases at $ T=0 $.
The solid line represents the critical line of
a first-order phase transition, 
the dashed line a second-order phase transition, 
and the dot-dashed line a crossover.
(b) The corresponding phase diagram in the $T$-$\rho_B$ plane 
in units of the nuclear matter density $\rho_{0}$.
There exist mixed phases corresponding to the first-order transitions
seen in (a).
}
\label{fig:pd_v00}
\end{center}
\end{figure}

As preliminary to the discussion on 
the effects of the vector interaction, 
we first present the phase structure without the  vector interaction.
We shall show that a coexisting phase appears where the quarks with 
dynamically generated mass are color superconducting.
This is a manifestation of competition between the
$\chi$SB and CSC phase transition.

In Fig.~\ref{fig:pd_v00}(a), 
the phase diagram in the $T$-$\mu$ plane is shown.
One can see  that 
there are  four different phases, i.e. the $\chi$SB phase,
the normal quark phase, which we call
the Wigner phase,
the CSC phase,
and a ``coexisting'' phase of $\chi$SB and CS; 
as seen from the upper small 
panel, which is an enlargement of the part around
the solid line near zero temperature, 
the last phase, in which quarks with dynamically generated
mass are color superconducting, occupies
only a small region in the $T$-$\mu$ plane 
near zero temperature with $\mu$ slightly 
smaller than $\mu_{B0}=1035$ MeV.

In the figure, 
the critical line of first- and second-order transitions
are represented by 
the solid and dashed lines, respectively;
notice that there exists a dashed line in the upper small panel.
We remark that there are three kinds of first-order transitions:
$\chi$SB-Wigner, $\chi$SB-CSC and coexisting-CSC transitions.
An artificial critical line of the crossover chiral transition 
is also shown  by the dash-dotted line 
on which the dynamical quark mass 
takes the same value as that at the endpoint $M_{D} = 186$ MeV, so 
that the crossover critical line is connected continuously 
with the critical line of the first-order transition
at the endpoint.
\footnote{
We have followed the criterion used 
in Ref.~\citen{ref:AY}. 
}
With this definition of the critical line for 
the crossover chiral transition, 
the critical temperature at vanishing chemical 
potential ($\mu=0$) is found to be $187$ MeV, which 
is slightly larger than the critical temperature
obtained in simulations of lattice QCD with 
two flavors\cite{ref:Lattice2}.

In accordance with a widely accepted view\cite{ref:RW-review,ref:A},
one sees that the chiral transition is first-order 
at low temperatures:
The critical line of the first-order transition emerging 
from a point in the zero-temperature line
terminates at
\begin{eqnarray}
(T_e,\mu_{Be})=(47,990)~{\rm MeV}. \nonumber 
\end{eqnarray}
The figure also shows that 
the phase transition from the CSC to the Wigner phase is second 
order when $T$ is raised in our model, in which the gluon fields are not
explicitly included\cite{ref:BL}.
We have found that 
the $\chi$SB phase is transformed into the coexisting 
phase at very low temperatures by a second-order transition
when $\mu$ is raised, as shown in the small panel.

\begin{figure}[t]
\begin{center}
\includegraphics[scale=1.3]{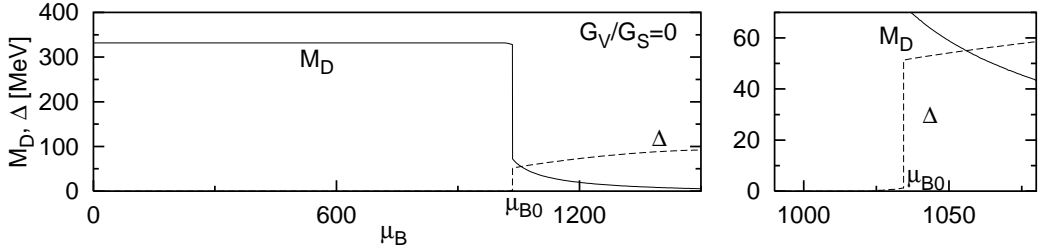}
\caption{
The order parameters $M_D$ and $\Delta$ at $T=0$ 
as functions of $\mu_B$
with $ G_V=0 $.
There are discontinuities of the order parameters at $\mu_{B0}=1035$ MeV.
An enlarged figure near the critical point is also shown.
The gap $\Delta$ is finite  even in the region $ \mu_B<\mu_{B0}$.
}
\label{fig:op_v00}
\end{center}
\end{figure}

To see more detail of the coexisting phase, 
we show the $\mu$ dependence of  $M_D$ and $\Delta$ 
at $T=0$ in Fig.~\ref{fig:op_v00}.
One  can see that 
$M_D$ ($\Delta$) shows 
a discontinuous decrease (increase) at $\mu_B=\mu_{B0}$, 
which clearly indicates a first-order chiral (CSC) transition 
at this point.
A notable point here is that  
$M_{D}$ has a finite value even in the CSC phase, 
because of the finite current quark mass;
notice that $M_D$ is proportional to the 
chiral condensate and not the total (constituent)
quark mass $M=m+M_D$.
Although we do not show the result here,
we have checked that 
$M_{D}$ vanishes in the CSC phase in the chiral limit 
(nevertheless see Fig. \ref{fig:op_cl_v20} for $G_V/G_S=0.2$).
On the other hand, 
there is a region  in which 
$\Delta$ becomes finite in the $\chi$SB phase,
 which remains the case in the chiral limit.
We have called this phase the coexisting phase.

We notice that  
a coexisting phase similar to ours was obtained
in some previous works.
\cite{ref:RSSV-coex,carter,mishra,pengfei,jackson1}
\footnote{
In Ref.~\citen{ref:SKP},
 the full coupled gap equations for $M_{D}$ and $\Delta$ 
were not solved, which is necessary 
to find the coexisting phase. 
The coexisting phase discussed in Ref.~\citen{abuki} 
using the NJL model is thermally unstable.
}
For example,
Rapp et al. \cite{ref:RSSV-coex} showed that 
the instanton-anti-instanton molecule model 
admits such a coexisting phase at finite chemical potential,
although the phase structure at $T\not=0$ was not examined.
However, they questioned the robustness of
the existence of the coexisting phase, because
other calculations using a similar NJL-type chiral model,\cite{ref:BR}
in which the effective scalar coupling constant $G_S$ in our notation 
is relatively large, did not exhibit such a coexisting phase.
In fact, we have also checked that if a slightly larger $G_S$ is used,
the coexisting phase disappears even in our case.
We shall  show, however, that the vector interaction induces
a competition between the $\chi$SB and CSC phase transition, 
and thereby the existence of the coexisting phase  
always becomes possible with a sufficiently large vector coupling.
\footnote{We remark that  
if the ratio $G_C/G_S$ in our notation is artificially large,
the coexisting phase can be also  realized in a broad region
in the $T$-$\mu$ plane,
as  shown in the random matrix model \cite{jackson1} and 
in the NJL model\cite{pengfei}.}

\begin{figure}[t]
\begin{center}
{\includegraphics[scale=1]{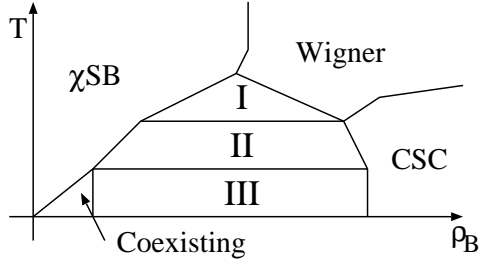}}
\caption{A schematic figure accounting for  Fig. \ref{fig:pd_v00}(b).
There are three mixed phases in the $T$-$\rho_B$ plane:
I is a mixed phase of the $\chi$SB and Wigner phases,
while II and III are mixed phases of the $\chi$SB and CSC phases,
and the coexisting and CSC phases, respectively.
}
\label{fig:mixed}
\end{center}
\end{figure}

The phase diagram in the $T$-$\rho_B$ plane is shown
in Fig.~\ref{fig:pd_v00}(b).
This phase structure is schematically presented 
in Fig.~\ref{fig:mixed}.
Corresponding to the three types of 
first-order transitions mentioned above,
there exist three mixed phases, which we call I, II and 
III, respectively:
I is a mixed phase of 
the $\chi$SB and Wigner phases, while 
II and III are mixed phases 
of the $\chi$SB and CSC phases, and the coexisting and CSC phases, 
respectively.
We remark that various other mixed phases are possible when 
the CSC phase is incorporated than when it is not.

This ends our investigation of 
the phase structure without the vector interaction.
When the vector interaction is included, 
the phase structure may be changed significantly, which 
we shall show is indeed the case in the next subsection.

\subsection{Phase structure with the vector interaction}

In this subsection, we discuss 
effects of the vector interaction on the phase structure 
of hot and dense quark matter by 
varying the vector coupling $G_{V}$ 
by hand in the range of $G_V/G_S=0 - 0.5$.
One will see that the vector interaction causes a nontrivial 
interplay between the $\chi$SB and CS phase, causing the optimal 
condensates to greatly fluctuate in a combined way.
This effect was not elucidated in the previous work\cite{ref:BHO}.

\begin{figure}
\begin{center}
\begin{tabular}{cc}
{\includegraphics[scale=1]{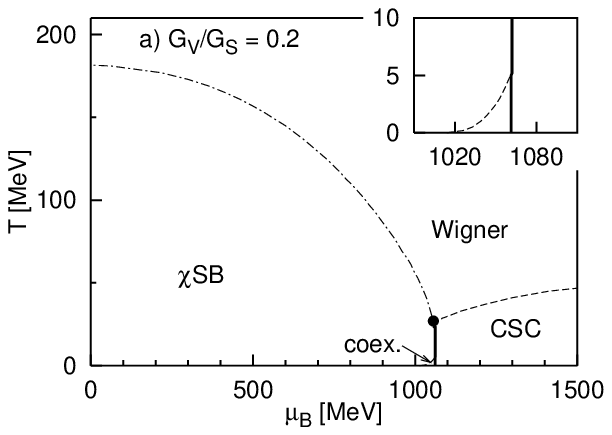}} & \hspace{-5mm}
{\includegraphics[scale=1]{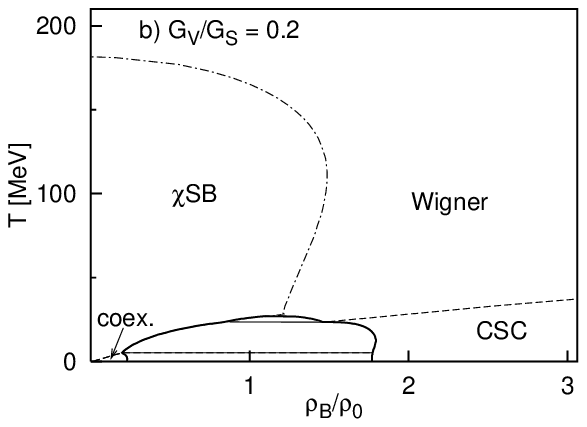}}
\end{tabular}
\caption{
The phase diagrams with  $ G_V/G_S=0.2 $ 
in the $T$-$\mu$ plane (a) and $T$-$\rho$ plane (b).
The solid line represents the critical line
of a first-order phase transition, 
the dashed line a second-order transition and the dot-dashed line a crossover.
}
\label{fig:pd_v20}
\end{center}
\end{figure}

The  phase structure in the $T$-$\mu$ plane
with $G_V/G_S=0.2$
is  shown  in Fig.~\ref{fig:pd_v20}(a).
The phase diagram consists of 
the $\chi$SB, Wigner, CSC and coexisting phases, 
as in Fig. \ref{fig:pd_v00}(a).
The dash-dotted line is the contour line at $M_D=198$ MeV
and is supposed to denote the critical line
of the crossover transition;
the solid and dashed lines represent the critical lines
of the first-order and second-order transitions, respectively,
as  in Fig.~\ref{fig:pd_v00}(a).
The corresponding phase diagram in the $T$-$\rho$ plane 
has the three mixed phases I, II and III which are 
seen in Fig.~\ref{fig:mixed},
as well as the $\chi$SB, Wigner and CSC phases.

\begin{figure}
\begin{center}
{\includegraphics[scale=1.3]{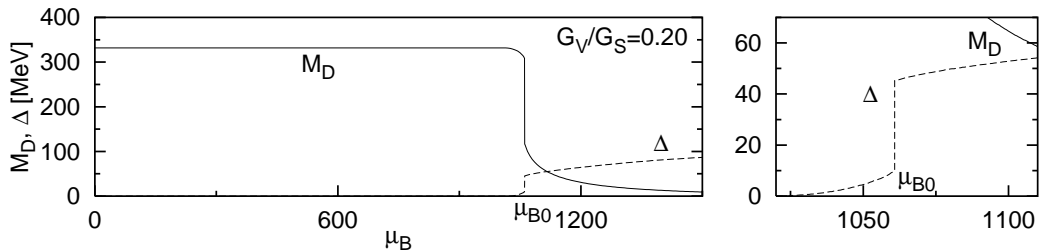}}
\caption{The order parameters $M_D$ and $\Delta$ 
 as a function of $\mu$ at $T=0$ with 
$G_V/G_S=0.2$.
}
\label{fig:op_v20}
\end{center}
\end{figure}

\begin{figure}
\begin{center}
{\includegraphics[scale=1.3]{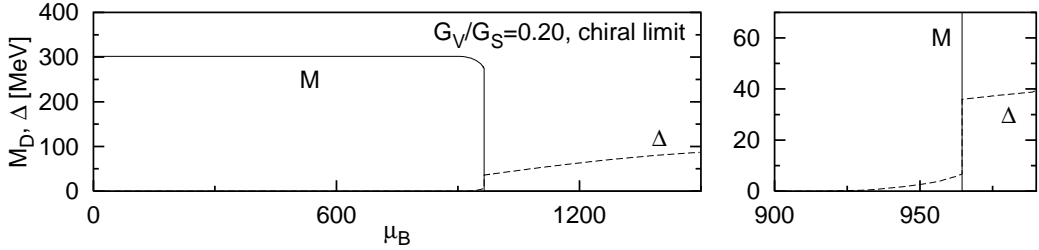}}
\caption{The order parameters $ M_D$ 
and $\Delta$ as functions of $\mu_B$ at $T=0$ in the chiral limit
with $G_V/G_S=0.2$.
The other parameters are slightly changed 
so as to reproduce the physical quantities in the chiral limit: 
$G_S=5.01\mbox~{GeV}^{-2}$, $\Lambda=650$ MeV
and $G_C=3.11\mbox{GeV}^{-2}$.
One can see that
the chiral condensate vanishes completely in the CSC phase while
it has a finite value in the coexisting phase.
Thus the phase transition
from and to the coexisting phase can be unambiguously defined.
}
\label{fig:op_cl_v20}
\end{center}
\end{figure}

From these figures, the following points are notable:\\ 
(1)~
The endpoint of the first-order transition moves 
toward a lower temperature and higher chemical potential,
\beq
(T_e,\mu_{Be})=(27,1056)~{\rm MeV}.\nonumber 
\eeq 
(2)~
The chiral restoration is moved toward larger $\mu$.
This is because 
the gap equations (\ref{eqn:2GEM}) and (\ref{eqn:2GED})
are functions of $T$ and $\tilde\mu$, 
and thus the explicit $G_V$ dependence is absorbed into $\tilde\mu$,
as shown in Eq. (\ref{shift}).
This means that $\mu$ given by a fixed $M_D$ 
is shifted toward larger values as $G_V$ is increased.\\
(3)~
The region of the coexisting phase becomes broader
in both $T$ and $\mu$ directions
in the $T$-$\mu$ plane, and hence also in the $T$-$\rho$ plane.
This feature is determined dominantly by the behavior 
of $M_{D}$ in the $\chi$SB phase.
As an example, $M_{D}$ together with 
$\Delta$, as a function of $\mu_B$ 
at $T=0$ is shown in Fig.~\ref{fig:op_v20};
the same quantities in the chiral limit are shown in
Fig.~\ref{fig:op_cl_v20}.
One sees that there appears a small region of $\mu_B$, smaller
than but near $\mu_{B0}$ in which 
$M_D$ ($\Delta$) shows a gradual decrease (increase); accordingly, 
finite $M_D$ and $\Delta$ coexist in this region.
One should notice here that although the coexistence in this sense 
is realized even when $\mu_B>\mu_{B0}$, as seen in Fig.~\ref{fig:op_v20},
$M_D$ in the chiral limit 
vanishes identically in this region, while the coexistence
of $M_D$ and $\Delta$ remains for 
at $\mu_B<\mu_{B0}$ as seen in Fig.~\ref{fig:op_cl_v20}.
In fact, this is also the case when $G_V=0$, as was noted in \S 4.1.
Thus, calling this the ``phase coexisting'' makes sense.
Anyway, the gradual change of the order parameters means that 
the first-order transition is weakened.
The decrease of $M_D$ also 
implies that of the total quark mass $M$, 
leading to a growth of the Fermi surface for a given $\mu_q$.
The larger the Fermi surface, the larger the gap $\Delta$, 
owing to the BCS mechanism. Thus
the region of the coexisting phase in the $T$-$\mu_B$ plane becomes broader.
This feature can be applied to the case for $T\neq 0$.
In short, the vector interaction promotes 
the formation of the coexisting phase.
This is one of the points which Buballa et al.\cite{ref:BHO}
did not clarified, because they used a fixed vector coupling.
\footnote{
The coexisting phase does not appear 
in Ref. \citen{ref:BHO}, although a relatively large  ratio
$ G_V/G_S=0.5 $ in our notation is adopted.
However, we should also notice 
that a larger $G_S$  leading to a 
larger constituent quark mass $M$ than ours is used there.
This suggests that 
if the driving force responsible for the $\chi$SB phase as represented by
$G_S$ is strong, a  larger ratio $G_V/G_S$ is needed
for  the realization of the coexisting phase.
It is worth mentioning in this respect that
the coexisting phase is obtained in the two-color QCD
on the lattice in a robust way\cite{ref:KTS2}.
}
It would be interesting to explore the possible correlation between the
appearance of the coexisting phase
and  the strength of the effective vector coupling extracted, 
say, 
from the baryon-number susceptibility,\cite{ref:Kuni2,ref:BR-rep,ref:HKRS} 
as was done in Ref.~\citen{ref:boyd}.

The characteristics (1) and (2) of the effects of the vector
interaction have been 
known to exist for the chiral transition without the CSC transition
incorporated.\cite{ref:AY,ref:KLW,ref:Bub}
(An account of the phase structure without the CSC transition 
is presented in Appendix \ref{sec:v4c} as a reference.)

\begin{figure}[t]
\begin{center}
\begin{tabular}{cc}
{\includegraphics[scale=1]{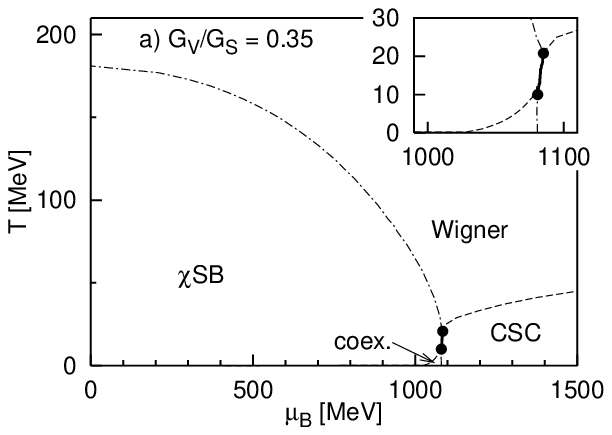}} & \hspace{-5mm}
{\includegraphics[scale=1]{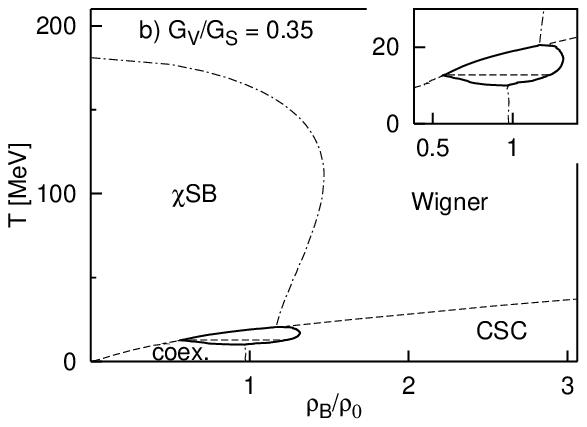}}
\end{tabular}
\caption{The phase diagram with $G_V/G_S=0.35$
in the $T$-$\mu$ plane (a) and $T$-$\rho$ plane (b).
There appear two endpoints of the first-order transition.
}
\label{fig:pd_v35}
\end{center}
\end{figure}

However, when the interplay between the $\chi$SB and CS phases 
enhanced with the vector interaction is taken into account, 
the variation of the phase diagram becomes  not so simple 
for larger $G_V$.
In Fig.~\ref{fig:pd_v35}(a), we show the phase diagram in 
the $T$-$\mu$ plane with $G_V/G_S=0.35$.
It is noteworthy that 
there appear two endpoints at both sides of the critical line 
of the first-order transition.
Accordingly, the coexisting-CSC transition at low temperatures 
becomes a crossover transition.
We have checked that 
the crossover transition becomes second order 
in the chiral limit, and hence a tricritical point 
appears instead of the endpoint of the first-order transition. 
As far as we know, 
this is the first time it has been shown that 
the critical line of the first-order transition
for the chiral restoration can have another endpoint
on the low temperature side, implying that the transition 
from the $\chi$SB phase to 
the CSC phase at low temperatures becomes a crossover 
(second order in the chiral limit).
Nevertheless it is noteworthy that 
the two-color QCD on the lattice at nonzero temperature and chemical
potential gives a similar phase diagram; 
see Fig.~1 of Ref.~\citen{ref:KTS2}.
Again, the lattice result might be interpreted in terms of the
effective vector coupling, which deserves exploration 
for the purpose of understanding the underlying physics.
The two-endpoint structure does not appear 
in the random matrix model with two colors\cite{jackson2},
in which, however, only the two auxiliary fields 
$\sigma \sim \langle \bar{\psi}\psi \rangle$ and 
$\Delta \sim \langle \psi C i\gamma_{5}\tau_{2}\tau_{2} \psi \rangle$ 
are explicitly introduced, but not with the vector field.
It would be intriguing to study whether 
such a phase structure can be realized
in the random matrix model with 
the incorporation of the vector field as an auxiliary field.

\begin{figure}[t]
\begin{center}
{\includegraphics[scale=1]{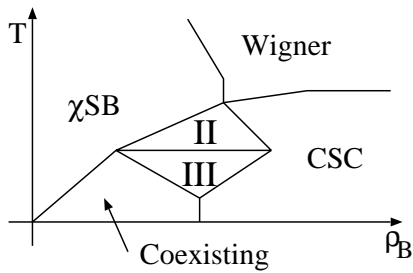}}
\caption{A schematic figure accounting for Fig. \ref{fig:pd_v35}(b).
The mixed phase I of the $\chi$SB and Wigner phases
does not exist in this case.
}
\label{fig:mixed2}
\end{center}
\end{figure}

The corresponding phase diagram in the $T$-$\rho$ plane 
is shown in Fig.~\ref{fig:pd_v35}(b).
Its schematic phase structure is represented in Fig.~\ref{fig:mixed2}.
The phases II and III correspond to 
the mixed phases of the $\chi$SB and CSC phases, 
and the coexisting and CSC phases, respectively as in 
Fig.~\ref{fig:mixed}.
Notably, the phase I does not exist anymore.

\begin{figure}
\begin{center}
\includegraphics[scale=1.1]{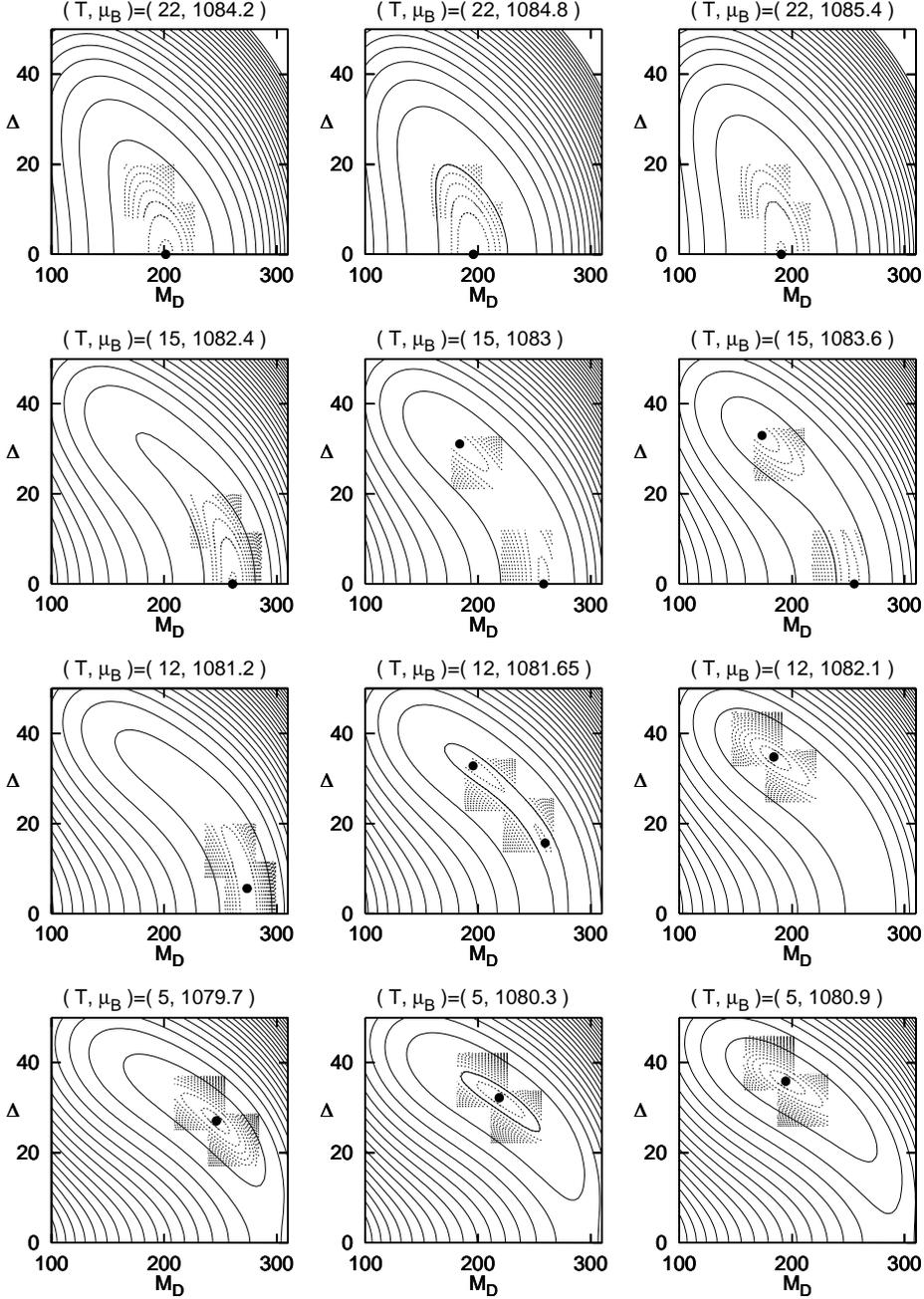}
\caption{
The contour of the thermodynamic potential
in the $M_D$-$\Delta$ plane for various values of $(T, \mu_B)$ 
around the critical point of the first-order transition.
The difference between the values of $\omega$ for 
adjacent solid lines 
is $1.5\times 10^6 \mbox~{MeV}^4$.
As shown in the second and third row panels,
there appear two local minima  at $T=12$ MeV and $T=15$ MeV
with $\mu_B$ near the critical value,
which indicates
 that the phase transition is first order at these temperatures.
On the other hand,
as shown in the bottom and top panels, 
there always exists only one local, and hence, the absolute minimum
at $T=5$ and $22$ MeV,
which minimum moves continuously as $\mu_B$ is increased,
implying that  the phase transition is a crossover.
}
\label{fig:ep35}
\end{center}
\end{figure}

To examine the mechanism  of the appearance of the two end 
points in detail, we show
the thermodynamic potentials in 
the $M_D$-$\Delta$ plane for various $T$ and $\mu$ in Fig.~\ref{fig:ep35}.
In the lowest panels, 
the thermodynamic potential at $T=5$ MeV is shown:
We see  only one local, and hence, the absolute minimum point, which 
varies continuously as $\mu$ is increased. This implies that 
the phase transition is a crossover at $T=5$ MeV.
At higher temperatures, however,
the thermodynamic potential  comes to have 
two local minima near the critical point, 
as  shown in the second and third row panels  for 
$T=12$ MeV and $T=15$ MeV, respectively,
and the phase transition becomes first order.
At even higher temperatures,
the double-minimum structure ceases to exist and 
the thermodynamic potential has only one local minimum again, 
as shown in the uppermost panel for  
$T=22$ MeV, and the phase transition again becomes a crossover.

In our model calculation, 
the two-end-point structure of the phase diagram
appears for finite $G_V$ but in a narrow range of $G_V/G_S$, i.e. 
$0.33 \lesssim G_V/G_S \lesssim 0.38$.
We should also note that even when the phase transition is 
first order, 
the height of the bump  
between the two local minima 
of the thermodynamic potential {\em per particle} 
is so small that it is found to be
comparable with or smaller than the temperature. 
This means that thermal fluctuations, which are ignored in the
mean-field approximation employed in this work, may easily
destroy the two-end-point structure.
What we have found is that the inclusion of the vector 
interaction makes the minimum of the thermodynamic potential
shallow in the $M_D$-$\Delta $ plane, suggesting the
significance of  fluctuations of the chiral and 
diquark condensates in a combined manner.
The incorporation of the thermal fluctuations is 
beyond the scope of this work.

\begin{figure}[t]
\begin{center}
\begin{tabular}{cc}
{\includegraphics[scale=1]{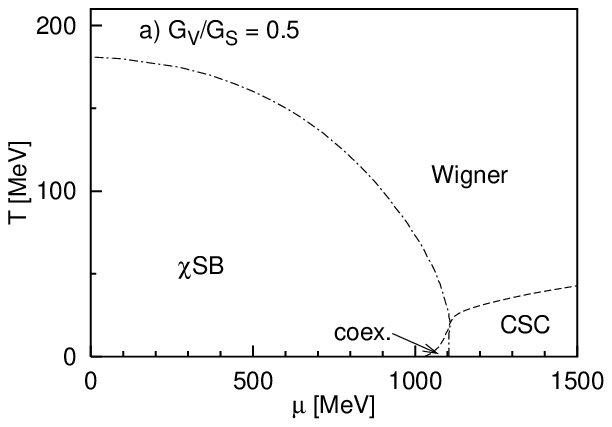}} & \hspace{-5mm}
{\includegraphics[scale=1]{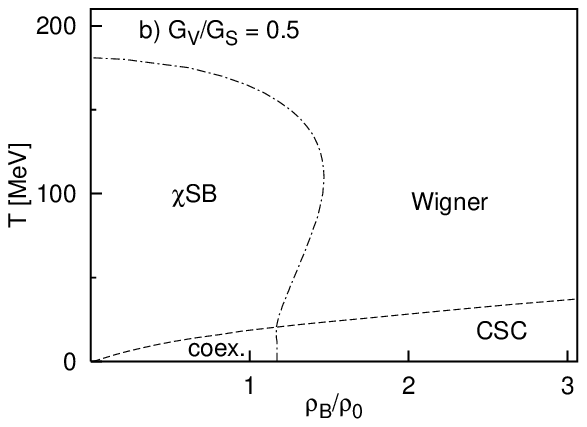}} 
\end{tabular}
\caption{(a) The phase diagram 
in the $T$-$\mu$ plane with $G_V/V_S=0.5$.
(b) The corresponding phase diagram in the $T$-$\rho$ plane.
}
\label{fig:pd_v50}
\end{center}
\end{figure}

When we consider a larger value of $G_V$ than $0.38G_S$, 
the first-order transition disappears
and it is changed completely into a crossover transition.
Figure \ref{fig:pd_v50}(a) shows the phase diagram in the $T$-$\mu$ plane 
with $G_V/G_S=0.5$ as a typical example in this case.
The dashed line denotes the second-order transition.
The dash-dotted line represents an artificial 
crossover line on which $M_D=200 $ MeV.
The corresponding phase diagram in the $T$-$\rho$ plane 
is shown in Fig.~\ref{fig:pd_v50}(b).
One can see that 
there is no first-order transition, and hence no mixed phase.
As pointed out in \S 3, 
the vector interaction affects the phase diagram in 
the $T$-$\rho$ plane 
only when there is a first-order transition.
Therefore, the phase structure in Fig.~\ref{fig:pd_v50} 
no longer changes after $G_V/G_S$ exceeds $0.38$.

\begin{figure}
\begin{center}
\begin{tabular}{cc}
{\includegraphics[scale=0.9]{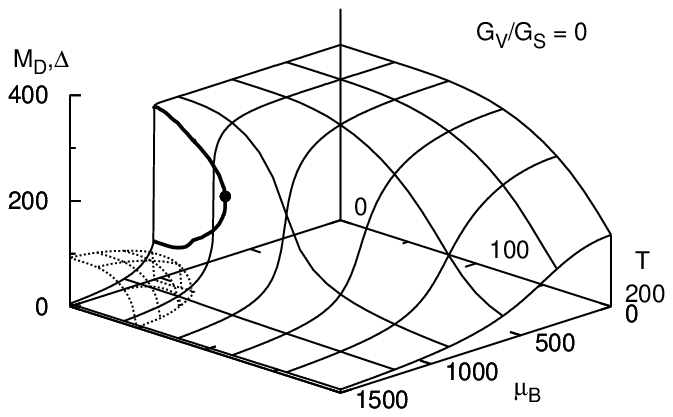}} &
{\includegraphics[scale=0.9]{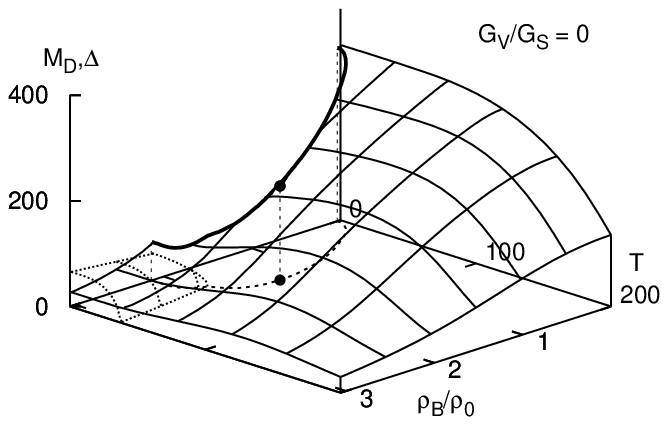}} \\
{\includegraphics[scale=0.9]{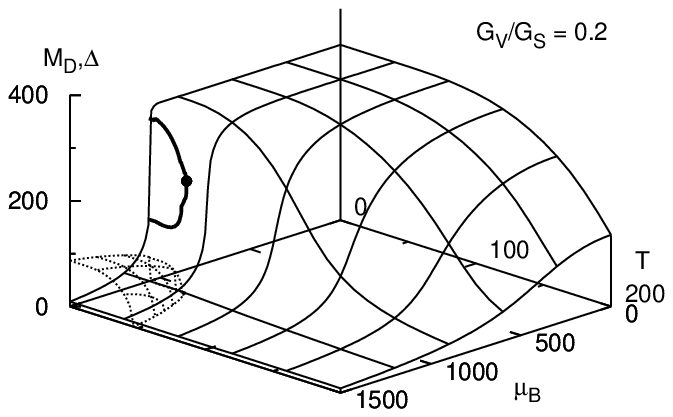}} &
{\includegraphics[scale=0.9]{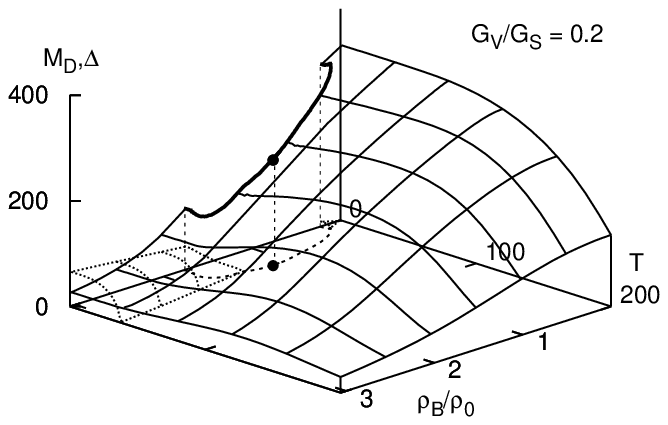}} \\
{\includegraphics[scale=0.9]{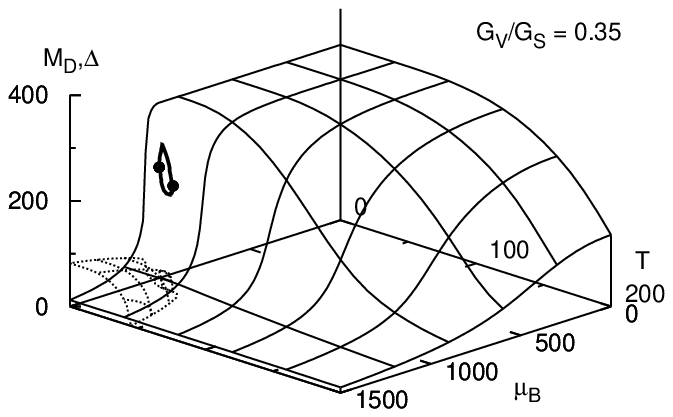}} &
{\includegraphics[scale=0.9]{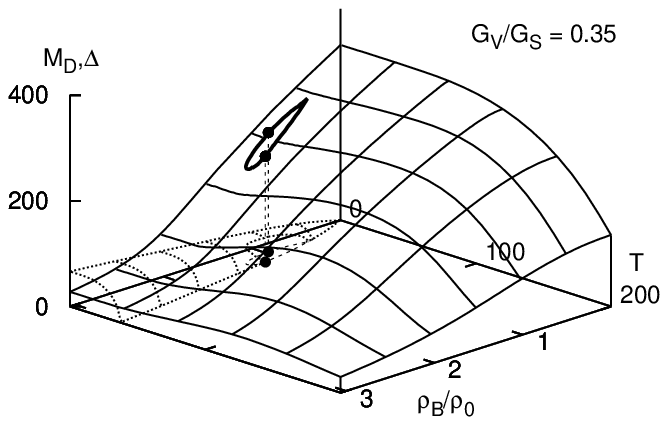}} \\
{\includegraphics[scale=0.9]{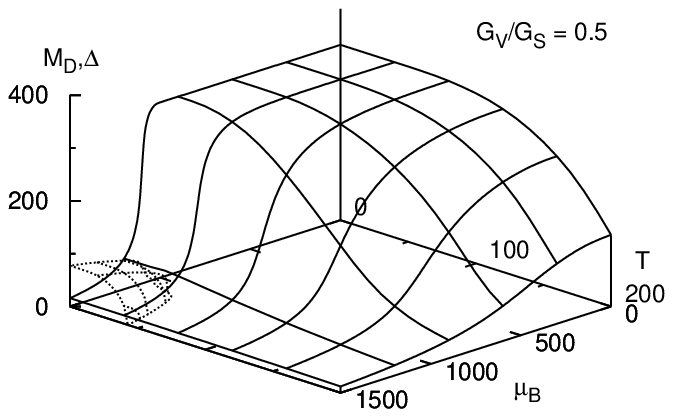}} &
{\includegraphics[scale=0.9]{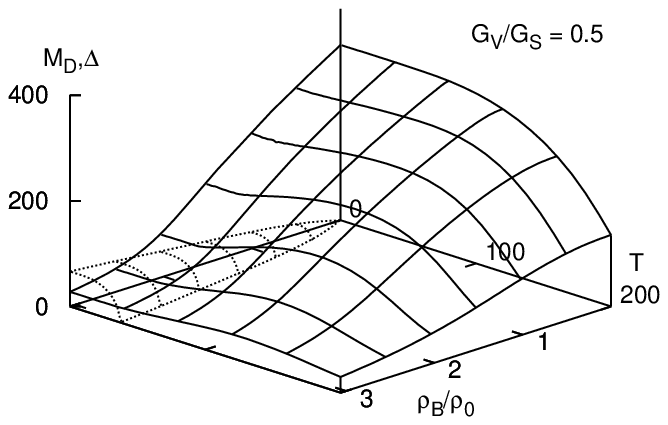}}
\end{tabular}
\caption{The left panels show
three-dimensional plots of the order parameters $M_D$ (solid lines) 
and $\Delta$ (dashed lines)
as functions of ($T, \mu_B$) with various values of $G_V/G_S$, while
the right panels show them as functions of ($T, \rho_B$).
The thick line corresponds to the first-order transition
and  the circles denote their endpoints.
Notice that the behavior of the order parameters
as functions of ($T, \rho_B$)
does not depend on $G_V$, except in the region of the mixed phases, 
in accordance with the discussion given in \S3.
}
\label{fig:opr_v50}
\end{center}
\end{figure}

As a nice summary of the effects of the vector interaction on the
phase structure of hot and/or dense quark matter,
we show three-dimensional plots of 
the dynamical quark mass $M_D$  
and the gap $\Delta$ in the $T$-$\mu$ and $T$-$\rho$ plane in 
Fig.~\ref{fig:opr_v50}.
The thick 
line represents the critical line of the first-order transition. 
The dotted points indicate the endpoints.
One sees that 
$M_D$ decreases more smoothly for larger $G_V$ in the $T$-$\mu$ plane.
It is clear that the $G_V$ dependence
of $M_D$ and $\Delta$ in the $T$-$\rho$ plane appears 
only in the critical region of the first-order transition.

\section{Summary and concluding remarks}

We have investigated  effects of the 
vector coupling on the chiral and color superconducting 
phase transitions at finite density and temperature
in a simple Nambu-Jona-Lasinio 
model by focusing on the implication of the density jump accompanied 
by the chiral transition.
We have shown that the phase structure is strongly 
affected by the vector interaction, 
especially near the critical line between the 
chiral symmetry breaking ($\chi$SB)
and color superconducting (CSC) phases: 
The first-order transition
between the $\chi$SB and CSC phases becomes weaker
as the vector coupling is increased,
and there can exist two endpoints of the critical line
of the first-order restoration in some range of parameters values;
the two endpoints become tricritical points in the chiral limit. 
Our calculation has shown that the repulsive vector interaction enhances
the competition  between the $\chi$SB and CS phases, 
leading to a degeneracy in the thermodynamic potential 
in the $M_D$-$\Delta$ plane.
This implies that there exists gigantic fluctuations 
of the order parameters that appear in a correlated way 
near the critical region, and it suggests the necessity of 
a theoretical  treatment incorporating the fluctuations.
This is, however, beyond the scope of this work.
We have found that the coexisting phase, in which the 
quarks with dynamically generated mass are 
color-superconducting, appears in a wide range of values 
of $\mu_B$ and $T$.
Here it should be emphasized that it is not yet known whether 
the chiral and the confinement-deconfinement transitions 
occur simultaneously at finite density; 
hence there may exist quark matter with chiral symmetry breaking.
We have emphasized that the appearance of such a coexistence
phase becomes robust and hence universal through the inclusion of 
the vector interaction.
We have also shown that the repulsive vector interaction
causes the transition from the chirally broken phase to color 
super conducting phase to move toward larger $\mu_B$.

Although our analysis is based on a simple model,
our finding that the vector interaction 
enhances the competition 
between the $\chi$SB and CSC phase transitions is universal and should
be confirmed and further studied 
with more realistic models, including the random matrix model 
and on lattice QCD.
In fact, phase structure similar to that found here has been obtained in 
two-color QCD on the lattice \cite{ref:KTS,ref:KTS2}, in which 
there appear two tricritical points related to the chiral and CSC
transitions, and also 
the coexisting phase in a wide range of the
temperature and chemical potential.
It may be possible to intuitively understand these results 
in terms of the effective vector coupling, which can be extracted by 
calculating the baryon-number susceptibility.
A random matrix study of the QCD phase diagram\cite{jackson1,jackson2} 
incorporating the vector condensate (i.e.,
the density) explicitly should be carried out.

In this work, we have ignored color neutrality 
in the CSC phase
\cite{ref:iida,ar-neutral,steiner,pegfei-neutral}.
It is known, however, that color neutrality
seems to have an only small effect on the onset of CS:
When the superconducting gap is sufficiently smaller 
than the Fermi momentum, as is the case 
in a wide region of the $T$-$\mu$ plane in our calculation, 
the densities of the paired and unpaired quarks 
are close in magnitude.
Therefore, the present results obtained 
for the effects of the vector coupling will only slightly change 
when the color neutrality is taken into account, 
although  the two-end-point structure, which is 
realized through a delicate interplay between $\chi$SB
and CS through the vector coupling,
might disappear or persist in the mean-field approximation
we have employed.

In the two-flavor case which we have treated in this paper,
the first two-color states which form the color Cooper 
pairs should have different dynamical mass from the 
remaining color state.
Although incorporating such color-dependent dynamical masses
\cite{ref:RSSV-coex,ref:BHO,ref:HT} is known to import 
only a tiny effect, it should be 
taken into account for a complete analysis of the effects of 
the vector coupling on the phase boundaries.

Furthermore, when applying the theory to neutron star phenomena,
the charge neutrality and the beta equilibrium condition incorporating
degenerate neutrinos\cite{ar-neutral,steiner} 
should also be taken into account.
We have confined our investigation to the two-flavor case in this work.
Needless to say, it would be very interesting to examine the effects of
the vector interaction in the three-flavor case, and thereby on the
color-flavor locked phase\cite{ref:CFL,ref:ABR,ref:BO,ref:GNA}.

T.Kunihiro thanks David Blaschke for informing him of the work
by Buballa et al.\cite{ref:BHO} and related papers after the completion 
of this work in May.
We are grateful to Michael Buballa for pointing out us some 
misleading statements in the original manuscript 
with regard to Ref. \citen{ref:BHO} and subsequent discussions
for elucidating the relation between  the present work and 
Ref.~\citen{ref:BHO}.
M. Kitazawa thanks 
H. Abuki for communications confirming the precise meaning 
of his master thesis\cite{abuki}.
This work is partially supported 
by Grants-in-Aid from 
the Japanese Ministry of Education, Science and Culture 
(No. 12640263 and 14540263).

\begin{figure}[t]
\begin{center}
\begin{tabular}{cc}
{\includegraphics[scale=1]{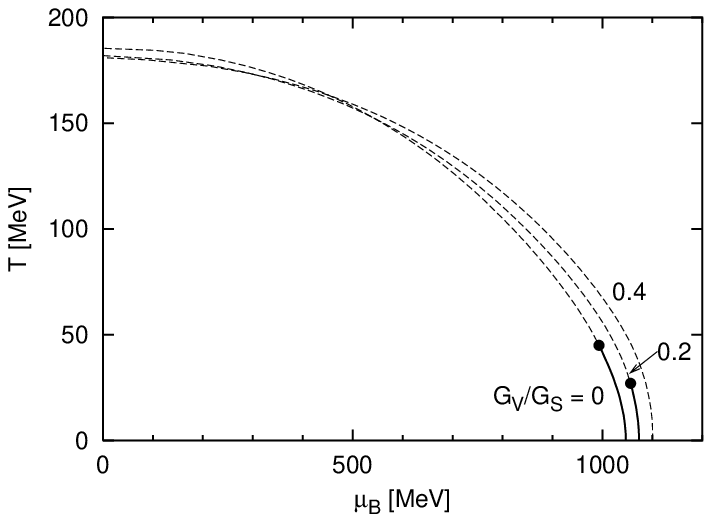}}& \hspace{-6mm}
{\includegraphics[scale=0.8]{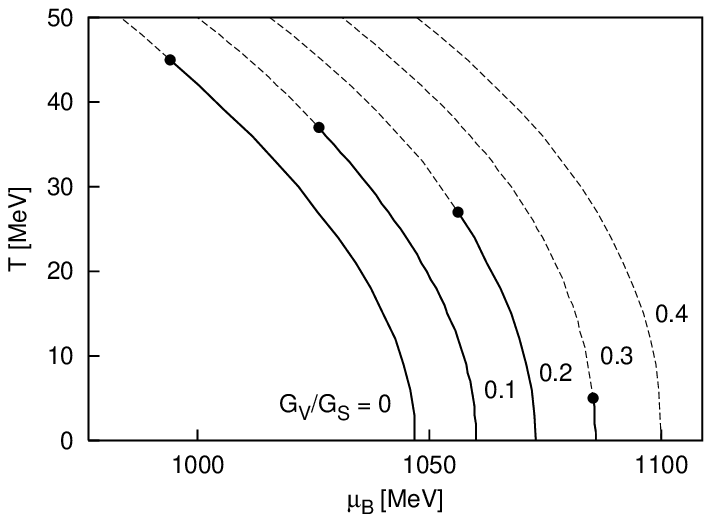}}
\end{tabular}
\caption{The $G_V$ dependence of 
the phase diagram for the chiral transition 
in the $T$-$\mu$ plane.
The solid line represents the critical line of the 
first-order transition.
The dash-dotted line denotes
an artificial critical line of the crossover transition, 
which is determined with the same condition as that stated in the text.
}
\label{fig:pdc}
\end{center}
\end{figure}

\appendix
\section{Effects of the Vector Interaction on the Chiral Phase Transition}
\label{sec:v4c}

In this appendix, we summarize 
how the chiral transition is  affected by the vector interaction 
in the case that the CS is not incorporated.
Although 
this problem has been examined by some authors,\cite{ref:Bub,ref:BR}
no coherent summary has been given in the literature.

\begin{figure}[t]
\begin{center}
{\includegraphics[scale=1]{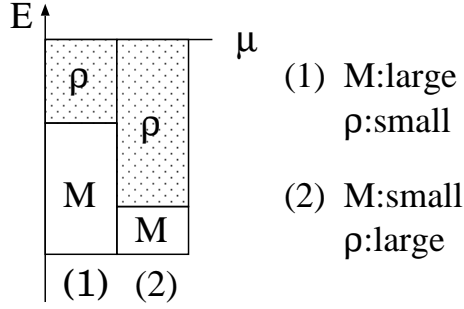}}
\caption{A conceptual diagram accounting for the relation between
the total  quark mass $M=m+M_D$
and the density $\rho_q$ at given $\mu_q$.
The quark density in the equilibrium state
becomes small for larger $M$ with $\mu_q$ fixed.
}
\label{fig:M-rho}
\end{center}
\end{figure}

The phase diagram of the chiral transition
in the $T$-$\mu$ plane is shown in Fig.~\ref{fig:pdc}.
Here we have used the same Lagrangian (\ref{eqn:NJL})
as that used in the
text, but with $G_C$ switched off.
One can see the following features from Fig.~\ref{fig:pdc}:\\
(i)~ The chiral restoration is shifted toward larger $\mu$
as $G_V$ is increased.\\
(ii)~ $G_V$ acts to moves the endpoint toward lower $T$ and larger 
$\mu$.\\ 
(iii)~
The chiral restoration  eventually turns into a 
crossover transition for large $G_V$.

\begin{figure}[t]
\begin{center}
{\includegraphics[scale=1]{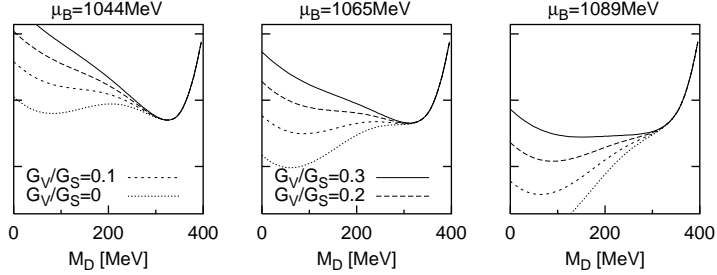}}
\caption{The thermodynamic potential at $T=0$ 
as a function of
$M_D$ for $G_V=0,0.1,0.2,0.3$.
For smaller $G_V$, $\omega$ has two local minima, reflecting a 
first-order transition.
As $G_V$ is increased, $\omega _D$ at small $M_D$ becomes 
large and the local minimum at smaller $M_D$ disappears.
In this way, the chiral transition becomes a crossover.}
\label{fig:efc}
\end{center}
\end{figure}

The feature (i) can be understood as follows.
The Fermi momentum $ p_F = \sqrt{ \mu^2-M^2 } $
becomes large (small) for small (large) $M$,
where $M=m+M_D$ is a constituent (total) quark mass,
and so does the density $\rho$ at the fixed $\mu$
(see Fig.~\ref{fig:M-rho}).
Since the vector interaction gives rise to a repulsive energy
proportional to the density squared, $ G_V\rho_q^2 $,
 a system with a smaller density is favored when $G_V$ is present.
Thus one can see when $G_V$ is finite,
the larger $M$ is favored.
We show the thermodynamic potential $\omega$ as a function of $M_D$
with various $G_V$ in Fig.~\ref{fig:efc}.
We see that the 
thermodynamic potential at small chiral condensate $M_D$
increases as $G_V$ increases, 
owing to the repulsion of the vector interaction.
Accordingly, the chiral restoration is shifted toward large $\mu$
as $G_V$ is increased.

Figure~\ref{fig:efc} also shows that
the first-order transition is weakened as $G_V$ is increased:
One sees from the far left panel 
that the thermodynamic potential with $G_V=0$
has two local minima and there exists a bump between these minima.
This two-local minima structure becomes less prominent and the
local minima becomes closer as $G_V$ is increased
(see the $G_V/G_S=0.2$ case (short-dashed line) in the middle panel).
Such two-local minima structure disappears at 
$G_V=0.3$ for all $\mu_B$.

\end{document}